\documentclass[aps,pra,preprint,groupedaddress,amsmath,amssymb,showpacs]{revtex4-1}
\usepackage{graphicx,amsmath,amssymb,epsfig}
\usepackage{dcolumn}
\usepackage{bm}
\begin{document}
\preprint{}

\title{Crossover from Electromagnetically Induced Transparency to Autler-Townes Splitting
in Open V-Type Molecular Systems}
\author{Chengjie Zhu}
\affiliation{State Key Laboratory of Precision Spectroscopy and Department of Physics,
East China Normal University, Shanghai 200062, China}

\author{Chaohua Tan}
\affiliation{State Key Laboratory of Precision Spectroscopy and Department of Physics,
East China Normal University, Shanghai 200062, China}

\author{Guoxiang Huang}
\email[Email: ] {gxhuang@phys.ecnu.edu.cn}

\affiliation{State Key Laboratory of Precision Spectroscopy and Department of Physics,
East China Normal University, Shanghai 200062, China}

\date{\today}

\begin{abstract}

We investigate electromagnetically induced transparency (EIT) and  Autler-Townes splitting (ATS) in an open V-type molecular system. Through detailed analytical calculations on the absorption spectrum of probe laser field by using residue theorem and spectrum decomposition, we find that EIT may occur and there exists a crossover from EIT to ATS (i.e. EIT-ATS crossover) for hot molecules. However, there is no EIT and hence no EIT-ATS crossover for cold molecules. Furthermore, we prove that for hot molecules EIT is allowed even for a counter-propagating configuration. We provide explicit formulas of EIT conditions and widths of EIT transparency windows of probe field when hot molecules work in co-propagating and counter-propagating configurations, respectively. Our theoretical result agrees well with the recent experimental one reported by Lazoudis {\it et al}. [Phys. Rev. A {\bf 83}, 063419 (2011)].

\end{abstract}


\maketitle

\section{INTRODUCTION}{\label{Sec:1}}

Quantum coherent phenomena occur widely in multi-level systems interacting resonantly with electro-magnetic fields. ln 1955, Autler and Townes \cite{at} showed that an absorption line of molecular transition can split into two Lorentzian lines (doublet) when one of two levels involved in the transition is coupled to a third one by a strong microwave field. Such doublet is now called Autler-Townes splitting (ATS). In 1961, Fano \cite{fano} showed that two resonant modes decaying via a common reservoir may yield a quantum destructive interference between the modes mediated by the reservoir. Such phenomenon is now called Fano interference.

In recent years, much attention has been paid to the study on electromagnetically induced transparency (EIT). By use of the quantum interference effect induced by a control field,  significant suppression of absorption of a probe field can be realized, together with large reduction of group velocity and giant enhancement of Kerr nonlinearity \cite{Fleischhauer2005}. Due to EIT, a transparency window appears in probe-field absorption spectrum, which can be generally decomposed into two Lorentzian terms, together with one (or several) Fano interference term(s). Thus EIT line shape displays characters of both ATS and Fano interference.

In the past two decades, EIT  and related quantum interference effects in various atomic systems has been studied intensively in both theory and experiment, and a large amount of research progress has been achieved \cite{Fleischhauer2005,Khur}. Similar phenomena in molecular systems have also been explored in recent years. Especially,  EIT has been observed in $^7$Li$_2$ \cite{Qi2002,Lazoudis2010}, K$_2$ \cite{Li2005} and Na$_2$ vapors \cite{Lazoudis2008,Lazoudis2011}, in  acetylene molecules filled in hollow-core photonic crystal fibers \cite{Ghosh2005,Benabid2011} and in photonic microcells \cite{Light2009}, and in Cs$_2$ in a vapor cell \cite{Li2010}, and so on.

Although many experiments have been carried out, up to now theoretical approach on EIT in molecular systems is less developed. Unlike atoms, even simplest molecules are open systems in which each excited molecular rovibrational level is radiatively coupled to many other energy levels. Furthermore, all related experiments were made by using thermal molecular vapors, which involve Doppler broadening and other decoherence effects. Therefore, EIT in molecular systems is more challenging not only for experimental observation, but also for theoretical analysis. Because of the
difficulty for analytical approach, numerical simulations are usually taken. However, the result of numerical simulations is generally not easy to clarify various EIT characters, and also hard to distinguish ATS from EIT clearly.

In the present work, we develop an analytical approach on EIT and ATS in an open V-type molecular system. By detailed analytical calculations on the absorption spectrum of probe field using residue theorem and spectrum-decomposition method, we find that EIT is possible and there exists a crossover from EIT to ATS (i.e. EIT-ATS crossover) for hot molecules with Doppler broadening. In contrast, there is no EIT and hence no EIT-ATS crossover for cold molecules.  We find also that for hot molecules EIT is allowed even for a counter-propagating configuration for probe and control fields. We provide explicit formulas of EIT conditions, widths of EIT transparency windows, and group velocities of probe field when hot molecules work in co-propagating and counter-propagating configurations, respectively. Our theoretical result agrees well with the recent experiment reported by Lazoudis {\it et al}. \cite{Lazoudis2011}.

Before proceeding, we notice that the spectrum-decomposition method was first proposed by Agarwal \cite{Agarwal1997} for analyzing probe-field absorption in several typical three-level atomic systems, which can isolate the precise nature of quantum interference induced by a control field. However, Agarwal's method is valid only for strong control field. Lately, Anisimov  and Kocharovskaya \cite{Petr2008} considered absorption lineshape of $\Lambda$-type system in view of resonant poles, and successfully explained the nature of quantum destructive interference for weak control field. Recently, Abi-Salloum \cite{Tony2010} distinguished EIT and ATS for similar atomic systems discussed in Ref. \cite{Agarwal1997}  by using the method in Ref.~\cite{Petr2008}, but with this method one cannot obtain the quantum interference term for strong control field analytically. In a recent work, Anisimov {\it et al}.  suggested a computational fitting technique to objectively discerning ATS from EIT \cite{Petr2011} from experimental data. Very recently, an experimental investigation of the crossover between ATS and EIT was carried out by Giner {\it et al.} \cite{giner} by using the method proposed by Anisimov {\it et al.} \cite{Petr2011} These works are significant, especially for clarifying the difference between EIT and ATS and some related concepts of quantum interference.

From the  works \cite{Agarwal1997,Petr2008,Tony2010,Petr2011} mentioned above, we can define
EIT as a quantum coherent phenomenon, where not only a transparency window is opened in probe-field absorption spectrum, but also a quantum {\it destructive} interference induced by control field should appear.
Notice that such definition of EIT is very general, in which the reason of quantum destructive interference is not specified. The  quantum destructive interference can be induced by different physical mechanisms in different systems, including the V-type system we consider below.

The present work is related to Refs. \cite{Agarwal1997,Petr2008,Tony2010,Petr2011}, and in particular to the experimental work in Ref. \cite{Lazoudis2011}. However,
systems considered in Refs. \cite{Agarwal1997,Petr2008,Tony2010,Petr2011} are only for cold atomic systems. It has been shown by the authors of Refs. \cite{Agarwal1997,Petr2008,Tony2010,Petr2011} that
an EIT is impossible for a cold V-type system because the quantum interference in such system is constructive. Our work is the first analytical approach to discern ATS and EIT in Doppler-broadened molecular systems. We shall explicitly show that a quantum destructive interference and hence EIT may occur in V-type molecular systems.
Furthermore, our analytical approach developed below is valid for arbitrary control field, and can demonstrate clearly the contribution of Doppler broadening, and various quantum interference characters (EIT, ATS, and EIT-ATS crossover) in a clear way.

The article is organized as follows. Section \ref{Sec:2} presents our theoretical model for the open V-type molecular system. Section \ref{Sec:hot_m} provides the solution of the Maxweel-Bloch (MB) equation and discusses the absorption and dispersion properties of probe field. The spectrum decomposition and EIT-ATS crossover is analyzed in detail and a comparison between our theoretical result with the experimental one by Lazoudis {\it et al}. \cite{Lazoudis2011} is given. Section \ref{Sec:cold_m} studies the linear absorption of the probe field in corresponding cold molecular system. Section \ref{Sec:4} discusses the roles of saturation and hole burning in the V-type system.
Finally, Section \ref{Sec:5} summarizes the main results obtained in our work.

\section{MODEL AND GENERAL SOLUTION}{\label{Sec:2}}

\subsection{The Model}

\begin{figure}
  \includegraphics[scale=0.7]{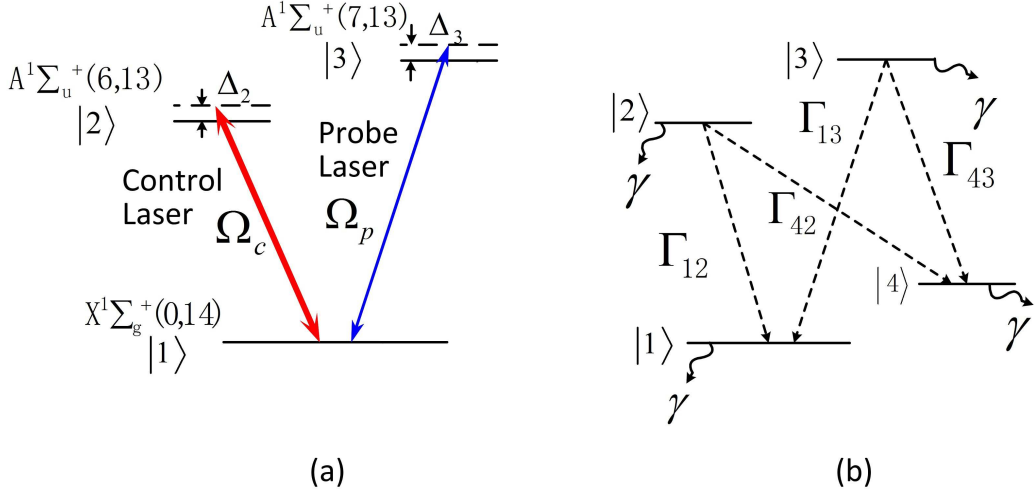}\\
  \caption{(Color online) (a): V-type three-level scheme of Na$_2$ molecular system. Ground state X$^1\Sigma_g^+(v^{''}=0,J^{''}=14)$ (labeled as $|1\rangle$) is coupled to the excited state A$^1\Sigma_u^+(v^{'}=6,J^{'}=13)$ (labeled as $|2\rangle$) by the control laser field with half Rabi frequency $\Omega_c$, and also to the excited state A$^1\Sigma_u^+(v^{'}=7,J^{'}=13)$ (labeled as $|3\rangle$) by the probe laser field with  half Rabi frequency $\Omega_p$. $\Delta_{2}$ and $\Delta_{3}$ are detunings of control and probe fields, respectively. (b): Molecule occupying the excited states $|2\rangle$ and $|3\rangle$ follow various relaxation pathways and decay to many ground-state levels besides the state $|1\rangle$. All these states are represented by the state $|4\rangle$. $\Gamma_{jl}$ denotes the spontaneous decay from state $|l\rangle$ to $|j\rangle$. $\gamma$ is the rate with which $\sigma_{jl}$ relaxes to its equilibrium value $\sigma_{jl}^{\rm eq}$.}\label{mod1}
\end{figure}

Our model is the same as that used in Ref. \cite{Lazoudis2011}. An open three-level  V-
type Na$_2$ molecular system (Fig.~\ref{mod1}(a))
%
%
consists of two exited upper molecular states A$^1\Sigma_u^+(v^{'}=6,J^{'}=13)$ (labeled $|2\rangle$),   A$^1\Sigma_u^+(v^{'}=7,J^{'}=13)$  (labeled $|3\rangle$),  and a ground state X$^1\Sigma_g^+(v^{''}=0,J^{''}=14)$ (labeled $|1\rangle$). A probe (control)  field with center frequency $\omega_p$ ($\omega_c$) and center wavevector ${\bf k}_p$ (${\bf k}_c$) couples to the excited state $|3\rangle$ ($|2\rangle$) and the ground state $|1\rangle$.  Electric field acting on the molecule system is of the form $\mathbf{E}=\sum_{l=p,c}\mathbf{e}_l{\cal E}_l(z,t)e^{i(\mathbf{k}_l\cdot\mathbf{r}-\omega_lt)}+$c.c., where $\mathbf{e}_l$ $({\cal E}_l)$ is the unit polarization vector (envelope) of $l$th electric-field component.
Both upper exited states ($|2\rangle$ and $|3\rangle$) are considered to decay spontaneously to the ground state $|1\rangle$ with decay rates $\Gamma_{12}$ and $\Gamma_{13}$, respectively. However, due to the open character of the system, the molecule occupying the excited states $|2\rangle$ and $|3\rangle$  may follow various relaxation pathways and decay to many ground states besides the state $|1\rangle$. For simplicity, all these states are represented by state $|4\rangle$. The decay rate $\Gamma_{4l}\ (l=2,3)$ indicates the spontaneous emission rate of level $|l\rangle$ to level $|4\rangle$ (see Fig.~\ref{mod1}(b)). The decay rate $\gamma$ is the rate with which $\sigma_{jl}$ relaxes to its thermodynamical equilibrium value $\sigma_{jl}^{\rm eq}$.

For hot molecules, Doppler broadening must be taken into account because the experiment are carried out in a heat-pipe oven \cite{Lazoudis2011}. Under electric-dipole and rotating-wave approximations, the interaction Hamiltonian of the
system in interaction picture reads
\begin{align}\label{H}
\hat{H}=-\hbar\left(\Omega_ce^{i[\mathbf{ k}_c\cdot(\mathbf{r}+\mathbf{v}t)-\omega_ct]}|2\rangle\langle1|+\Omega_pe^{i[\mathbf{k}_p\cdot(\mathbf{r}
+\mathbf{v}t)-\omega_pt]}|3\rangle\langle1|+{\rm c.c.}\right),
\end{align}
where $\bf{v}$ is molecular velocity,
$\Omega_{c(p)}=\left( {\bf e}_{c(p)}\cdot\boldsymbol{\mu}_{21(31)}\right) {\cal E}_{c(p)}/(2\hbar)$ is half Rabi frequency of the probe (control) field, with $\boldsymbol{\mu}_{jl}$ the electric-dipole matrix element associated with the transition from state $|j\rangle$ to state $|l\rangle$. Density matrix elements
in the interaction picture are $\sigma_{jl}=\rho_{jl}\exp{\{i[(\mathbf{k}_l-\mathbf{k}_j)\cdot(\mathbf{r}+\mathbf{v}t)
-((E_l-E_j)/\hbar-\Delta_l
+\Delta_j)t]\}}$ ($j,l$=1-4), here $\Delta_1=0$, $\Delta_2=\omega_c-(E_2-E_1)/\hbar$, and
$\Delta_3=\omega_p-(E_3-E_1)/\hbar$ are detunings and $\rho_{jl}$ is the density matrix elements in
Schr\"{o}dinger picture, with $E_j$ being the eigenenergy of the level $|j\rangle$.
Bloch equation governing the evolution of $\sigma_{jl}$ reads
\begin{subequations} \label{dme}
\begin{align}
&i\frac{\partial}{\partial t}\sigma_{11}+i\gamma(\sigma_{11}-\sigma_{11}^{\rm eq})-i\Gamma_{12}\sigma_{22}-i\Gamma_{13}\sigma_{33}+\Omega_c^\ast\sigma_{21}-\Omega_c\sigma_{21}^\ast
+\Omega_p^\ast\sigma_{31}-\Omega_p\sigma_{31}^\ast=0,\\
&i\frac{\partial}{\partial t}\sigma_{22}+i\gamma(\sigma_{22}-\sigma_{22}^{\rm eq})+i\Gamma_{2}\sigma_{22}+\Omega_c\sigma_{21}^\ast
-\Omega_c^\ast\sigma_{21}=0,\\
&i\frac{\partial}{\partial t}\sigma_{33}+i\gamma(\sigma_{33}-\sigma_{33}^{\rm eq})+i\Gamma_{3}\sigma_{33}+\Omega_p\sigma_{31}^\ast
-\Omega_p^\ast\sigma_{31}=0,\\
&i\frac{\partial}{\partial t}\sigma_{44}+i\gamma(\sigma_{44}-\sigma_{44}^{\rm eq})-i\Gamma_{42}\sigma_{22}-i\Gamma_{43}\sigma_{33}=0,\\
&i\frac{\partial}{\partial t}\sigma_{21}+d_{21}\sigma_{21}+\Omega_c(\sigma_{11}-\sigma_{22})-\Omega_p\sigma_{32}^\ast=0,\\
&i\frac{\partial}{\partial t}\sigma_{31}+d_{31}\sigma_{31}+\Omega_p(\sigma_{11}-\sigma_{33})-\Omega_c\sigma_{32}=0,\\
&i\frac{\partial}{\partial t}\sigma_{32}+d_{32}\sigma_{32}+\Omega_p\sigma_{21}^\ast-\Omega_c^\ast\sigma_{31}=0,
\end{align}
\end{subequations}
where $d_{21}=-\mathbf{k}_c\cdot\mathbf{v}+\Delta_{2}-\Delta_{1}+i\gamma_{21}$, $d_{31}=-\mathbf{k}_p\cdot\mathbf{v}+\Delta_{3}-\Delta_{1}+i\gamma_{31}$, $d_{32}=-(\mathbf{k}_p-\mathbf{k}_c)\cdot\mathbf{v}+\Delta_{3}-\Delta_{2}+i\gamma_{32}$ with $\gamma_{jl}=(\Gamma_{j}+\Gamma_{l})/2+\gamma+\gamma_{jl}^{\rm col}$ ($j,l$=1-3).
$\Gamma_{l}=\sum_{j\neq l}\Gamma_{jl}$ with $\Gamma_{jl}$ denoting the rate per molecule
at which population decays from state $|l\rangle$ to state $|j\rangle$.
Quantity $\gamma_{jl}^{\rm col}$ is the dephasing rate due to processes such as
elastic collisions. $\sigma_{jj}^{\rm eq}$ is the thermal equilibrium value of $\sigma_{jj}$
when all electric-fields are absent.  Equation (\ref{dme})  satisfies
$\sum_{j=1}^4 \sigma_{jj}=1$ with  $\sum_{j=1}^4 \sigma_{jj}^{\rm eq}=1$.
At thermal equilibrium, population in the excited states is much smaller than
that of the ground state, i.e. $\sigma_{22}^{\rm eq}=\sigma_{33}^{\rm eq}\simeq0$ and hence $\sigma_{11}^{\rm eq}+\sigma_{44}^{\rm eq}=1$.

Evolution of the electric field is governed by the Maxwell
equation
$\nabla^2 {\bf E}-(1/c^2)\partial^2{\bf E}/\partial
t^2=[1/(\epsilon_0 c^2)]\partial^2 {\bf P}/\partial t^2$,
with the electric polarization intensity given by
\begin{equation}
{\bf P}=
{\cal N}_a\int^{+\infty}_{-\infty} dvf(v) \left[ \boldsymbol{\mu}_{13}\sigma_{31} e^{i(k_p \cdot z-\omega_p
t)} +\boldsymbol{\mu}_{12}\sigma_{21} e^{i(k_c\cdot z-\omega_c
t)}+{\rm c.c.} \right],
\end{equation}
where ${\cal N}_a$ is molecular density and $f(v)$ is molecular velocity distribution function.
For simplicity, we have assumed electric-field wavevectors are along $z$-direction, i.e. ${\bf k}_{p,c}=(0,0,k_{p,c})$.
Under slowly-varying envelope approximation, the Maxwell equation reduces to
\begin{equation}\label{eqs:maxwell}
i\left(\frac{\partial}{\partial z}+\frac{1}{c}\frac{\partial}{\partial t}\right)\Omega_p
+\kappa_{13}\int^{+\infty}_{-\infty}dvf(v)\sigma_{31}=0,
\end{equation}
where $\kappa_{13}={\cal N}_a\omega_p|\boldsymbol{\mu}_{31}|^2/(2\hbar\varepsilon_0
c)$, with $c$ the light speed in vacuum.

\subsection{General solution}

The base state of the system, i.e. the time-independent solution of the Maxwell-Bloch
(MB) Eqs. (\ref{dme}) and (\ref{eqs:maxwell}) in the absence of the probe field is
\begin{subequations}\label{eqs:init}
\begin{eqnarray}
& & \sigma_{11}^{(0)}=\gamma(\gamma+\Gamma_3)\left(\gamma+\Gamma_2
+2\gamma_{21}\frac{|\Omega_c|^2}{|d_{21}|^2}\right)
\frac{\sigma_{11}^{\rm eq}}{D_0},\\
& & \sigma_{22}^{(0)}=2\gamma\gamma_{21} (\gamma+\Gamma_3)\frac{|\Omega_c|^2}{|d_{21}|^2}\,
\frac{\sigma_{11}^{\rm eq}}{D_0},\\
& &
\sigma_{21}^{(0)}=\frac{\Omega_c}{d_{21}}(\sigma_{22}^{(0)}-\sigma_{11}^{(0)})
=-\frac{\Omega_c}{d_{21}}\gamma(\gamma+\Gamma_2)(\gamma+\Gamma_3)\frac{\sigma_{11}^{\rm
eq}}{D_0},
\end{eqnarray}
\end{subequations}
and $\sigma^{(0)}_{33}=\sigma^{(0)}_{31}=\sigma^{(0)}_{32}=0$, where
$D_0=2\gamma_{21}(|\Omega_c|^2/|d_{21}|^2)
(\gamma+\Gamma_{42})(\gamma+\Gamma_{3})
+\gamma(\gamma+\Gamma_3)[\gamma+\Gamma_2+2\gamma_{21}(|\Omega_c|^2/|d_{21}|^2)]$.
Note that in above expressions $d_{21}=d_{21}(v)=-k_c
v+\Delta_{2}-\Delta_{1}+i\gamma_{21}$, $d_{31}=d_{31}(v)=-k_p
v+\Delta_{3}-\Delta_{1}+i\gamma_{31}$, and
$d_{32}=d_{32}(v)=-(k_p-k_c)v+\Delta_{3}-\Delta_{2}+i\gamma_{32}$.
Notice that $\sigma_{44}^{(0)}=1$ and all other
$\sigma_{ij}^{(0)}=0$ if $\gamma=0$. However, in our thermal molecular system
$\gamma\neq 0$ ($\gamma\approx 3$ MHz in the experiment \cite{Lazoudis2011}), thus
$\sigma^{(0)}_{ll}$ $(l=1,2)$
and $\sigma^{(0)}_{21}$ take non-zero values. In particular, for large
$\Omega_c$  molecules populate mainly in $|1\rangle$
and $|2\rangle$, i.e. the population in $|4\rangle$ is small (around 10\% of the total
number of the molecules).


At first order in $\Omega_p$, the populations and the coherence between the states $|1\rangle$ and
$|2\rangle$ are not changed. It is easy to get the solution
\begin{eqnarray}\label{1osolution}
&&\Omega_p^{(1)}=F\,e^{i[K(\omega) z-\omega t]}\\
&&\sigma_{31}^{(1)}=\frac{d_{21}^\ast(\omega+d_{32})\sigma_{11}^{(0)}
-|\Omega_c|^2(\sigma_{11}^{(0)}-\sigma_{22}^{(0)})}{d_{21}^\ast D}F\,e^{i[K(\omega) z-\omega t]}\\
&&\sigma_{32}^{(1)}=\frac{-(\omega+d_{31})\Omega_c^\ast(\sigma_{11}^{(0)}-\sigma_{22}^{(0)})
+d_{21}^\ast\Omega_c^\ast\sigma_{11}^{(0)}
}{d_{21}^\ast D}F\,e^{i[K(\omega) z-\omega t]}
\end{eqnarray}
where $D=|\Omega_c|^2-(\omega+d_{31})(\omega+d_{32})$, $F$ is a constant. Dispersion relation
$K(\omega)$ \cite{note2} reads
\begin{equation}\label{eq:Dp}
K(\omega)=\frac{\omega}{c}+\kappa_{13}\int^\infty_\infty dv f(v)\frac{d_{21}^\ast(\omega+d_{32})\sigma_{11}^{(0)}
-|\Omega_c|^2(\sigma_{11}^{(0)}-\sigma_{22}^{(0)})}{d_{21}^\ast[|\Omega_c|^2-(\omega
+d_{31})(\omega+d_{32})]}.
\end{equation}

Notice that the integrand in the dispersion relation (\ref{eq:Dp}) depends on
three factors. The first is the AC Stark effect induced by the control field, reflected in the denominator, corresponding to the appearance of dressed states out of the states $|1\rangle$ and $|2\rangle$, by which two Lorentzian peaks in probe-field absorption spectrum are shifted from their original positions. The second is, reflected in the numerator, proportional to $\sigma_{11}^{(0)}-\sigma_{22}^{(0)}$. The appearance of nonzero $\sigma_{22}^{(0)}$ is due to
the saturation effect induced by the control field.  When the control field grows, the saturation effect increases. When  $|\Omega_c|^2/|d_{21}|^2\gg 1$, $\sigma_{22}^{(0)}\approx \sigma_{11}^{(0)}$  and hence the second term in the numerator will disappear.  The third is the Doppler effect, reflected by $d_{jl}=d_{jl}(v)$ and the molecular velocity distribution $f(v)$, which may increases or decreases probe-field absorption, as shown below.

\section{Hot molecules}\label{Sec:hot_m}

For a thermal gas, the integration in Eq.~(\ref{eq:Dp}) over molecular velocity $v$
must carried out properly. In thermal equilibrium, the velocity distribution function is Maxwellian
$f(v)=[1/(\sqrt{\pi}\,v_T)]\exp{\left(-v^2/v_T^2\right)}$,
where $v_T=\sqrt{2k_BT/M}$ is most probable speed, and $M$ is molecular mass.
Because the integration in Eq.~(\ref{eq:Dp}) with the Maxwellian distribution
leads to a particular combination of error functions, which is very inconvenient
for a systematic analytical approach. As did in Refs.~\cite{jav,Lee2003},
we adopt the Lorentzian velocity distribution profile
$f(v)=v_T/[\sqrt{\pi}(v_T^2+v^2)]$ to replace the Maxwellian distribution.

We are interested in two different cases: co-propagating (i.e. $k_p\approx k_c$) and counter-propagating
(i.e. $k_p \approx-k_c$), which will be discussed separately.

\subsection{Co-propagating configuration}\label{Sec:Co}

\begin{figure}
  \includegraphics[scale=0.8]{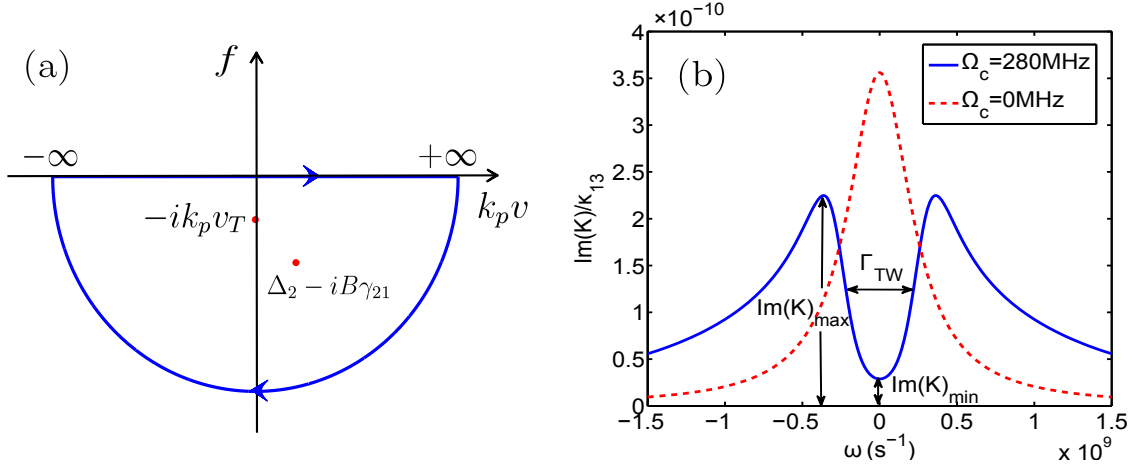}\\
  \caption{(Color online)
  (a): Two poles $(\Delta_2 ,-iB \gamma_{21} )$
  and $(0,-ik_pv_T)$ (represented by solid points) in the lower half complex plane of the
  integrand in Eq.~(\ref{eq:Dp}). The closed curve with arrows is the contour
  chosen for calculating the integration in Eq.~(\ref{eq:Dp}) by using residue
  theorem.
  (b): Probe-field absorption spectrum Im($K$) as a function of $\omega$.
  The solid (dashed) line for $|\Omega_c|=280$ MHz ($|\Omega_c|=0$ MHz). Definitions of Im$(K)_{\rm min}$, Im$(K)_{\rm max}$ and the width of transparency window $\Gamma_{\rm TW}$ are indicated in the figure.}\label{fig:res}
\end{figure}

In this case, one has $d_{21}=-k_pv+\Delta_2+i\gamma_{21}$, $d_{31}=-k_pv+\Delta_3+i\gamma_{31}$ and $d_{32}=\Delta_3-\Delta_2+i\gamma_{32}$. The second term on the right-hand side of Eq.~(\ref{eq:Dp})
can be calculated by considering a contour integration (see Fig. \ref{fig:res}(a)\,)
%
%
in a complex plane of $k_pv$ and using residue theorem \cite{bf}.

We find two poles in the lower half complex plane for the integrand of Eq.~(\ref{eq:Dp})
\begin{equation} \label{poles}
k_pv=\Delta_2-iB\gamma_{21},\,\,\,\,k_pv=-ik_pv_T,
\end{equation}
with
$B=\left\{ 1+[2|\Omega_c|^2/\gamma_{21}(\gamma+\Gamma_2)]\left[1
+(\gamma+\Gamma_{42})/\gamma \right]\right\}^{1/2}$.
By taking a contour consisting of real axis and a semi-circle in
the lower half complex plane (see the curves with arrows shown in Fig.~\ref{fig:res}(a)),
we can calculate the integration in Eq.~(\ref{eq:Dp}) analytically
by just calculating the residues corresponding to the two poles,
and obtain exact and explicit result for the integration. Since general expression
is lengthy, here we just write down the one with $\Delta_2=\Delta_3=0$ and
the condition $\Delta\omega_D\gg\gamma_{jl},\ \gamma$:
\begin{subequations}\label{K1}
\begin{eqnarray}
& & K(\omega)=\omega/c+{\cal K}_1+{\cal K}_2, \\
& & {\cal K}_1=\kappa_{13}'\frac{(\omega+i\gamma_{32})[\Gamma_3(1-B^2)\gamma_{31}
+2|\Omega_c|^2]-i\Gamma_3|\Omega_c|^2(1+B)}{\gamma\Gamma_3B(\Delta\omega_D^2
-B^2\gamma_{31}^2)[|\Omega_c|^2-(\omega+i\gamma_{31}+iB\gamma_{31})(\omega+i\gamma_{32})]},\\
& & {\cal K}_2=\kappa_{13}' \frac{(\omega+i\gamma_{32})[\Gamma_3(\gamma_{31}^2-\Delta\omega_D^2)
+2\gamma_{31}|\Omega_c|^2]-i\Gamma_3|\Omega_c|^2(\gamma_{31}
+\Delta\omega_D)}{[\gamma\Gamma_3(\gamma_{31}^2-\Delta\omega_D^2)
+2\gamma_{31}|\Omega_c|^2\Gamma_{13}][|\Omega_c|^2
-(\omega+i\gamma_{31}+i\Delta\omega_D)(\omega+i\gamma_{32})]},
\end{eqnarray}
\end{subequations}
where $\kappa_{13}'=\sqrt{\pi}\kappa_{13}\gamma\sigma_{11}^{\rm eq}$ and
$\Delta\omega_D=k_pv_T$ (Doppler width). Notice that for cold molecules
the second pole in Eq. (\ref{poles}) is absent, thus ${\cal K}_2=0$. However,
for hot molecules, due to Doppler effect one has ${\cal K}_2\neq 0$, and hence
the system may display very different quantum interference characters that do not
exist for cold molecules.

In most cases, $K(\omega)$ can be Taylor expanded around the center frequency
(corresponding to $\omega=0$) of the probe field, i.e., $K(\omega)
=K_0+K_1\omega+\dots$, where $K_j \equiv (\partial^jK/\partial\omega^j)_{\omega=0}$. Here
Re($K_0$) and Im($K_0$) describe respectively phase shift and absorption
per unit length, and Re($1/K_1$) ($\equiv v_g$) gives
group velocity  of the probe field.

\subsubsection{Transparency window in the absorption spectrum}

Shown in Fig.~\ref{fig:res}(b)
is the probe-field absorption spectrum Im($K$) as a function of $\omega$. The solid (dashed) line is for $|\Omega_c|=280$ MHz ($|\Omega_c|=0$), with other parameters given by $\Gamma_{j2} \approx \Gamma_{j3}\,\,(j=1,4)=4.08\times10^7$ Hz, $\gamma_{32}^{\rm col}\approx \gamma_{21}^{\rm col}\approx \gamma_{31}^{\rm col}= 5\times10^6$ Hz, $\gamma=3\times10^6$ Hz and $\Delta\omega_D=0.6$ GHz.
One sees that the absorption spectrum for $|\Omega_c|=0$ has only a single peak (the dashed line). However, a significant transparency window is opened for a non-zero $\Omega_c$ (the solid line).
The minimum (Im$(K)_{\rm min}$), maximum (Im$(K)_{\rm max}$),  and width of transparency window ($\Gamma_{\rm TW}$) have been defined in the figure.

From Eq. (\ref{K1}), we get
\begin{equation}\label{eq:Imk0}
{\rm Im}(K)_{\rm min}\approx\frac{\kappa_{13}'}{\Delta\omega_D}\frac{x+z/2+z\sqrt{x}/2}{(1+\sqrt{x})(x+z/2)(1+2\sqrt{x}/z)}
\end{equation}
where $x\equiv |\Omega_c|^2\gamma_{31}/(\gamma_{32}\Delta\omega_D^2)$,
$z\equiv \gamma_{31}/(\Delta\omega_D)$
are two dimensionless parameters. At the temperature in the experiment carried out in
Ref.~\cite{Lazoudis2011}, one has
$z\ll1$ because $\gamma_{31}\ll\Delta\omega_D$. From Eq. (\ref{eq:Imk0})
we obtain the following conclusions:
\begin{itemize}
\item For a large $|\Omega_c|$, i.e. $|\Omega_c|\gg \sqrt{\gamma_{32}/\gamma_{31}}\Delta \omega_{D}$
      (and hence $x\gg 1$),  Im($K)_{\rm min}$ is vanishing small;
\item For a small $|\Omega_c|$, i.e. $|\Omega_c|\ll \sqrt{\gamma_{32}/\gamma_{31}}\Delta \omega_{D}$
      but with $|\Omega_c|\approx \sqrt{\gamma_{31}\gamma_{32}}$
      (and hence $x\ll 1$, $\sqrt{x}/z\sim1$), Im($K)_{\rm min}$ is small.
\end{itemize}

The first conclusion is obvious because the reduction of Im($K)_{\rm min}$ for larger $|\Omega_c|$ is due to ATS effect. The maximum of Im$(K)$ is found to be Im$(K)_{\rm max}\approx \kappa_{13}'/\Delta\omega_D$, located at at $\omega\approx\pm\Omega_c$. Using these results we obtain the expression of the
width of the transparency window
\begin{equation}\label{Tr}
\Gamma_{{\rm TW}}\approx2\left(\frac{2|\Omega_c|^2+\Delta\omega_D^2-\Delta\omega_D\sqrt{\Delta\omega_D^2
+4|\Omega_c|^2}}{2}\right)^{1/2}.
\end{equation}

However, the second conclusion is not easy to understand because,
from conventional viewpoint,  for a small $|\Omega_c|$ the Doppler broadening suppresses the quantum interference effect induced by $\Omega_c$. In the following, we shall show that such conclusion is not correct for the V-type system with Doppler effect. In fact, the suppression of Im($K)_{\rm min}$ for small $|\Omega_c|$ can be obtained because the Doppler effect can contribute a quantum {\it destructive} interference to the system.

\subsubsection{EIT-ATS crossover}

Now we extend the spectrum-decomposition method introduced in Refs.~\cite{Agarwal1997,Petr2008,Tony2010,Petr2011}
to analyze the detailed characters of the probe-field absorption explicitly.
${\cal K}_j$ $(j=1,2)$ in Eq. (\ref{K1}) can be easily decomposed as
\begin{equation}\label{decom1}
{\cal K}_j=\alpha_j \left( \frac{A_{j+}}{\omega-\delta_{j+}}+\frac{A_{j-}}{\omega-\delta_{j-}}
\right),
\end{equation}
where $\alpha_j$, $A_{j\pm}$ are constants, $\delta_{j+}$ and $\delta_{j-}$ are two spectrum poles of
${\cal K}_j$, all of which have been given explicitly in Appendix~\ref{app:co}.

Our aim is to analyze the quantum interference effect, for which the expression of Im$({\cal K}_j)$ is required. However, their general expressions are long and complicated. In order to
illustrate the quantum interference effect in a clear way,
we decompose Im$({\cal K}_j)$ according to different regions of $\Omega_c$ as the following.

(i). {\it Weak control field region}  (i.e. $\Omega_c<\Omega_{\rm ref}\equiv \Delta\omega_D/2$):
Using similar approach by Anisimov {\it et al}. \cite{Petr2008,Petr2011}, we obtain
the imaginary part of ${\cal K}_{j}$ in this region as
\begin{equation}\label{eq:Form1_EIT}
{\rm Im}({\cal K}_{j})=\alpha_{j}\left(\frac{C_{j+}}{\omega^2+W_{j+}^2}
+\frac{C_{j-}}{\omega^2+W_{j-}^2}\right)
\end{equation}
with real constants
\begin{subequations}\label{Wpm}
\begin{eqnarray}
& & C_{j+}=W_{j+}(W_{j+}-\Gamma^w_j)/(W_{j+}-W_{j-}),\\
& & C_{j-}=-W_{j-}(W_{j-}-\Gamma^w_j)/(W_{j+}-W_{j-}),\\
& & W_{1\pm}=\frac{1}{2}\left[\gamma_{31}(1+B)+\gamma_{32}\pm\sqrt{[\gamma_{31}(1+B)
-\gamma_{32}]^2-4|\Omega_c|^2}\right],\\
& & W_{2\pm}=\frac{1}{2}\left[\gamma_{31}+\Delta\omega_D+\gamma_{32}\pm\sqrt{(\gamma_{31}
+\Delta\omega_D-\gamma_{32})^2-4|\Omega_c|^2}\right],
\end{eqnarray}
\end{subequations}
where
$\Gamma^w_1=\gamma_{32}-\Gamma_3|\Omega_c|^2(1+B)/[\Gamma_3(1-B^2)\gamma_{31}
+2|\Omega_c|^2]$ and $\Gamma^w_2=\gamma_{32}-\Gamma_3|\Omega_c|^2(\gamma_{31}
+\Delta\omega_D)/[\Gamma_3(\gamma_{31}^2-\Delta\omega_D^2)+2\gamma_{31}|\Omega_c|^2]$.

\begin{figure}
\includegraphics[scale=0.75]{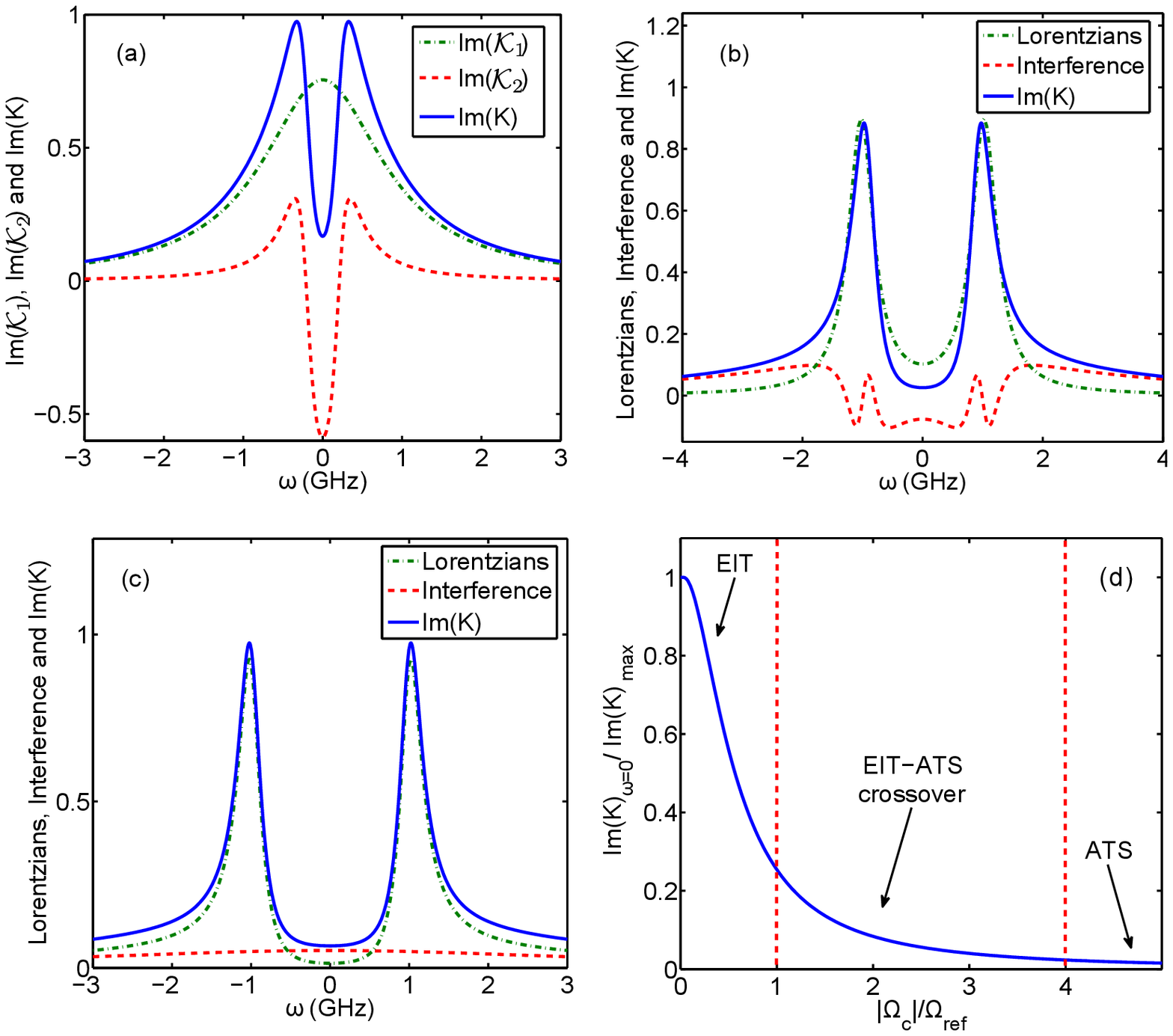}
\caption{(Color online)
EIT-ATS crossover for hot molecules in the co-propagating
configuration. (a):  Im(${\cal K}_1$) (dash-dotted line), Im(${\cal
K}_2$) (dashed line), and  ${\rm Im}({\cal K})$ (solid line) as
function of $\omega$ for $\Omega_c<\Omega_{\rm ref}\equiv
\Delta\omega_D/2$. (b): Absorption spectrum contributed by two
Lorentzians (dash-dotted line), the destructive interference (dashed
line), and total absorption spectrum Im($K$) (solid line), in the
region $\Omega_c>\Omega_{\rm ref}$. (c): Absorption spectrum
contributed by two Lorentzians (dash-dotted line), the small
constructive interference (dashed line), and total absorption
spectrum Im($K$) (solid line), in the region $\Omega_c\gg\Omega_{\rm
ref}$. Panels (a), (b) and (c) correspond to EIT, EIT-ATS crossover,
and ATS regions, respectively. (d): Transition from EIT to ATS for
hot molecules in the co-propagating configuration. Shown is ${\rm
Im}(K)_{\omega=0}/{\rm Im}(K)_{\rm max}$ as a function of
$|\Omega_c|/\Omega_{\rm ref}$. Three regions (EIT, EIT-ATS
crossover, and ATS) are divided by two dashed lines.} \label{L_D_co}
\end{figure}

Shown in Fig.~\ref{L_D_co}(a)
are results of Im(${\cal K}_1$) (the dash-dotted line),
Im(${\cal K}_2$) (the dashed line), and  Im(${K}$) (the solid line).
System parameters are given by $\Gamma_{j2} \approx \Gamma_{j3}\,\,(j=1,4)=4.08\times10^7$ Hz, $\gamma_{32}^{\rm col}\approx \gamma_{21}^{\rm col}\approx \gamma_{31}^{\rm col}= 5\times10^6$ Hz, $\gamma=3\times10^6$ Hz,
$\Omega_c=220$ MHz, and $\Delta \omega_D=0.6$ GHz.
We see that Im(${\cal K}_1$) is positive but Im(${\cal K}_2$) is negative. However, their
superposition, which gives Im(${K}$), is positive and displays a absorption doublet
with a significant transparency window near at $\omega=0$. Because there exists a destructive
interference in the probe-field absorption spectrum, the phenomenon found here should be attributed to
an EIT according to the criterion given in Refs.~\cite{Petr2008,Tony2010,Petr2011}.
Such EIT can be taken as the one induced by the Doppler effect. The reason is that when the Doppler broadening is absent, the negative Im(${\cal K}_2)$ term does not exist, and hence only an absorption spectrum with a positive single peak (i.e. the dash-dotted line contributed by Im(${\cal K}_1)$\,) appears.

(ii). {\it Large control field region} (i.e. $\Omega_c>\Omega_{\rm ref}$):
By extending the approach by Agarwal \cite{Agarwal1997},
we can decompose Im$({\cal K}_j)$ ($j=1,2$) as
\begin{eqnarray}\label{eq:Form2_EIT}
{\rm Im}({\cal K}_j)=\alpha_{j}& &\left\{\frac{1}{2}\left[\frac{W_j}{(\omega-\delta_{j}^r)^2+W_j^2}
+\frac{W_j}{(\omega+\delta_{j}^r)^2+W_j^2}\right]\right.\nonumber\\
& &\left.+\frac{g_j}{2\delta_{j}^r}\left[\frac{\omega-\delta_j^r}{(\omega-\delta_j^r)^2
+W_j^2}-\frac{\omega+\delta_j^r}{(\omega+\delta_j^r)^2+W_j^2}\right]\right\},
\end{eqnarray}
where
\begin{subequations}\label{eq:A2_co2}
\begin{eqnarray}
& & W_1=\frac{1}{2}[\gamma_{31}(1+B)+\gamma_{32}]/2, \\
& & \delta_1^r=\frac{1}{2}\left[4|\Omega_c|^2-[\gamma_{31}(1+B)-\gamma_{32}]^2\right]^{1/2}, \\
& & g_1=\frac{\gamma_{31}(1+B)-\gamma_{32}}{2}+\frac{\Gamma_3|\Omega_c|^2(1+B)}{
\Gamma_3(1-B^2)\gamma_{31}+2|\Omega_c|^2},
\end{eqnarray}
\end{subequations}
and
\begin{subequations}\label{eq:A1_co2}
\begin{eqnarray}
& & W_2=\frac{1}{2}(\gamma_{31}+\Delta\omega_D+\gamma_{32}), \\
& & \delta_2^r=\frac{1}{2}\left[4|\Omega_c|^2-(\gamma_{31}+\Delta\omega_D-\gamma_{32})^2\right], \\
& & g_2=\frac{\gamma_{31}+\Delta\omega_D-\gamma_{32}}{2}+\frac{\Gamma_3|\Omega_c|^2(\gamma_{31}
+\Delta\omega_D)}{\Gamma_3(\gamma_{31}^2-\Delta\omega_D^2)+2\gamma_{31}|\Omega_c|^2}.
\end{eqnarray}
\end{subequations}

Obviously, terms in the first square bracket on the right hand side of Eq. (\ref{eq:Form2_EIT})
are two Lorentzians, which are the net contribution to probe-field absorption from two different channels corresponding to the two dressed states created by the control field $\Omega_c$,
with  $W_j$ being the width (also strength) of the two Lorentzians and  $\delta_j^r$ being the real part of the spectrum poles. The following terms in the second square bracket are clearly quantum interference ones (which are called dispersive terms by Agarwal \cite{Agarwal1997}) . Obviously, the magnitude of the interference is controlled by the parameter $g_j$. If $g_j>0$ ($g_j<0$),
the interference is destructive (constructive).

Shown in Fig.~\ref{L_D_co}(b) are results of the probe-field absorption as functions of $\omega$ for
$\Omega_c>\Omega_{\rm ref}$. The dash-dotted line (dashed line) denotes the contribution by the sum of the Lorentzians terms (interference terms) in Im(${K}$) (=Im(${\cal K}_1$)+Im(${\cal K}_2$)\,). We see that the interference is destructive and, interestingly,  some structures appear. The solid line gives the result of Im($K$). System parameters used in the plot are the same as those in Fig.~3(a) but with $\Omega_c=1$ GHz. Clearly, a wide and deep transparency window is opened and the phenomenon found can be attributed to EIT-ATS crossover.

(iii). {\it Strong control field region} (i.e. $\Omega_c\gg\Omega_{\rm ref}$): In this region, the quantum interference strength $g_j/\delta_j^r$ in Eq.~(\ref{eq:Form2_EIT}) is very weak and negligible.
We have
\begin{equation}\label{eq:A1_co3}
{\rm Im}({\cal K}_j)\approx \frac{\alpha_j}{2}\left(\frac{W_j}{(\omega-\delta_j^r)^2+W_j^2}
+\frac{W_j}{(\omega+\delta_j^r)^2+W_j^2}\right),
\end{equation}
being to a sum of two equal-width Lorentzians shifted from the origin by $\delta_j^r\approx\pm\Omega_c$ ($j=1,2$).

Shown in Fig.~\ref{L_D_co}(c)  are results of the probe-field absorption as functions of $\omega$ for $\Omega_c \gg \Omega_{\rm ref}$. The dash-dotted line represents the contribution by the sum of the two Lorentzian terms. For illustration, we have also plotted the contribution from the interference terms (neglected in Eq. (\ref{eq:A1_co3})\,), denoted by the dashed line. We see that the interference is constructive but very small. The solid line is the curve of Im($K$), which has two resonances at $\omega\approx\pm\Omega_c$.
Parameters used are the same as those in panel (a) and (b)  but with $\Omega_c=1.8$ GHz. Obviously, the phenomenon found in this situation belongs to ATS  because the transparency window opened in this case is mainly due to the contribution of the
two Lorenztians.

The above results show that the probe-field absorption
spectrum experiences  a transition from EIT to ATS as the
control-field Rabi frequency $\Omega_c$ is changed from weak to
strong values. Essentially, one can obtain three different regions of the
probe absorption spectrum according to the value of
$|\Omega_c|/\Omega_{\rm ref}$. The first is EIT region
($|\Omega_c|/\Omega_{\rm ref}\leq 1$), where the quantum destructive
interference by the Doppler effect  results in the appearance of
transparency window. The second  is the one of EIT-ATS crossover
($1\leq|\Omega_c|/\Omega_{\rm ref}\leq4$), where both quantum
destructive interference and ATS exist together. Note that we have
defined ${\rm Im}(K)_{\omega=0}/{\rm Im}(K)_{\rm max}=0.01$ as the
border between EIT-ATS crossover and ATS regions. The third is ATS
region ($|\Omega_c|/\Omega_{\rm ref}>4$), where ${\rm
Im}(K)_{\omega=0}/{\rm Im}(K)_{\rm max}\leq0.01$ and the
transparency window is contributed only by the two Lorentzians.
Fig.~\ref{L_D_co} (d) shows a ``phase diagram'' that
%
%
illustrates the transition from the EIT to ATS  by plotting
${\rm Im}(K)_{\omega=0}/{\rm Im}(K)_{\rm max}$ as a function of $|\Omega_c|/\Omega_{\rm ref}$.

\subsubsection{Comparison with experiment}

\begin{figure}
\includegraphics[scale=0.7]{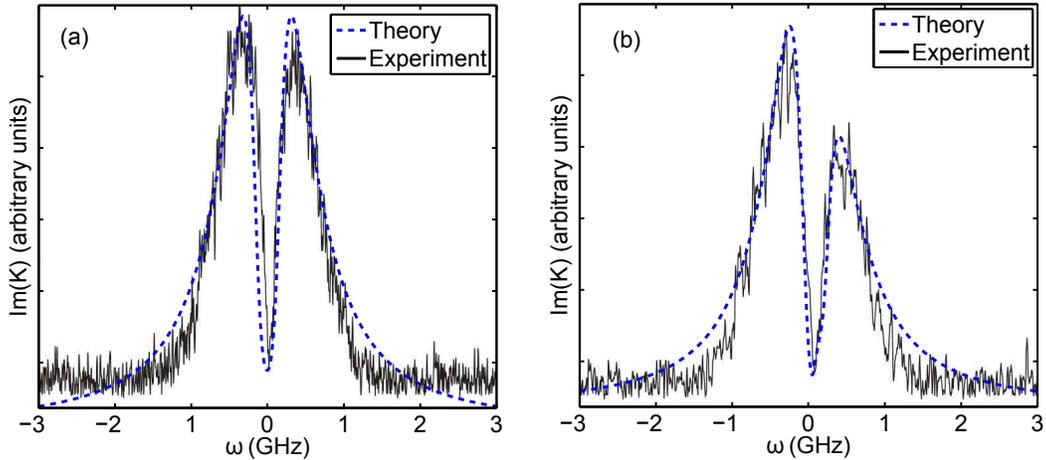}
\caption{(Color online) The experimental result  reported by  Lazoudis {\it et al}  \cite{Lazoudis2011} (solid line) and our theoretical one (dashed line) for the probe absorption spectrum in the case of the co-propagating configuration. (a): The control field is on resonance with $\Omega_c=220$ MHz. (b): The control field is detuned 100 MHz with $\Omega_c=190$ MHz.}\label{V_co}
\end{figure}

To verify our theoretical result given above, it is necessary to make a quantitative comparison with
the experimental one reported recently by Lazoudis {\it et al}  \cite{Lazoudis2011}. By using the parameters $\Gamma_{12}=\Gamma_{13}=\Gamma_{42}=\Gamma_{43}=4.08\times10^7$ Hz, $\gamma=3{\rm MHz}$, $\gamma_{jl}^{\rm col}\approx 5{\rm MHz}$, and $\Delta\omega_D=1.2$ GHz,
we calculate the absorption spectrum  ${\rm Im}(K)$ for the case of co-propagating configuration,
with results plotted as the dashed lines of  Fig.~\ref{V_co}.
%
%
The panel(a) of the figure is for the control field on resonance with $\Omega_c=220$ MHz, where a sharp dip appears in the center of the absorption spectrum and absorption doublet is symmetric. The panel (b) of the figure is for the control field detuned 100 MHz with $\Omega_c=190$ MHz, where a sharp dip also occurs but the absorption doublet is asymmetric.
One can see that our theoretical results (dashed lines) are very closed to the experimental ones measured by Lazoudis {\it et al}  \cite{Lazoudis2011}, which are represented by the solid lines. Note in passing that here we have plotted the quantity ${\rm Im}K$, which is proportional to fluorescence intensity related to the state $|3\rangle$ because $\sigma_{33}=2|\Omega_p|^2{\rm Im}(K)/(\gamma+\Gamma_3)$.
According to Eq.~(\ref{Tr}), the width of the transparency window $\Gamma_{TW}$, which is calculated to be $0.24$ GHz, agrees well with the experimental  one reported in Ref.~\cite{Lazoudis2011}. We stress that the system is in the region of weak control field (i.e. $\Omega_c\ll \Omega_{\rm ref}=\Delta\omega_D/2$), so the phenomenon observed by Lazoudis {\it et al}. \cite{Lazoudis2011} is
indeed an EIT phenomenon assisted by the Doppler effect.

\subsection{Counter-propagating configuration}

In this case, one has $d_{31}=\Delta_3-k_pv+i\gamma_{31}$,  $d_{21}=\Delta_2+k_pv+i\gamma_{21}$ and
$d_{32}=\Delta_3-\Delta_2-2k_pv+i\gamma_{32}$. Similarly, one can obtain the dispersion relation
with the form
\begin{subequations}\label{eq:CounterK}
\begin{eqnarray}
K=& & \frac{\omega}{c}+{\cal K}_1(\omega)+{\cal K}_2(\omega),\\
{\cal K}_1=& & \kappa_{13}'\frac{(\omega+i\gamma_{32})\Delta\omega_D[\Gamma_3(1-B^2)\gamma_{31}+2|\Omega_c|^2]
-i\Delta\omega_D\Gamma_3|\Omega_c|^2(1-B)}{\gamma\Gamma_3B(\Delta\omega_D^2-B^2\gamma_{31}^2)[|\Omega_c|^2
-(\omega+iB\gamma_{31}+i\gamma_{31})(\omega+i2B\gamma_{31}+i\gamma_{32})]},\\
{\cal K}_2=& & \kappa_{13}'\frac{(\omega+i\gamma_{32}+i2\Delta\omega_D)[(\gamma_{31}^2-\Delta\omega_D^2)+|\Omega_c|^2]
-i|\Omega_c|^2(\gamma_{31}-\Delta\omega_D)}{[\gamma(\gamma_{31}^2-\Delta\omega_D^2)+|\Omega_c|^2\Gamma_{13}][|\Omega_c|^2
-(\omega+i\gamma_{31}+i\Delta\omega_D)(\omega+i\gamma_{32}+i2\Delta\omega_D)]},
\end{eqnarray}
\end{subequations}
where ${\cal K}_1$ and ${\cal K}_2$ are obtained from poles $k_pv=\Delta_2-iB\gamma_{21}$
and $k_pv=-ik_pv_T$, respectively.

We first discuss the minimum value of the absorption spectrum at
$\omega=0$, i.e. ${\rm Im}(K)_{\rm min}$. From Eq. (\ref{eq:CounterK}), we obtain
\begin{equation}\label{eq:Imk0'}
{\rm Im}(K)_{\rm min}\approx\frac{\kappa_{13}'}{\Delta\omega_D}\frac{x+z/2+z\sqrt{x}/2+2\sqrt{x}/z}{(1+\sqrt{x})
(x+\sqrt{x}/z+z/2)(1+2\sqrt{x}/z)}
\end{equation}
where
$x\equiv |\Omega_c|^2\gamma_{31}/(2\gamma_{32}\Delta\omega_D^2)$ and $z\equiv \gamma/\Delta\omega_D$.
Obviously, we have $z\ll1$ because $\gamma\ll\Delta\omega_D$. From Eq. (\ref{eq:Imk0'})
we obtain the following conclusions:
\begin{itemize}
\item For a large $|\Omega_c|$, i.e. $|\Omega_c|\gg \sqrt{2\gamma_{32}/\gamma_{31}}\Delta \omega_{D}$
      (and hence $x\gg 1$),  Im($K)_{\rm min}$ is vanishing small;
\item For a small $|\Omega_c|$, i.e. $|\Omega_c|\ll \sqrt{2\gamma_{32}/\gamma_{31}}\Delta \omega_{D}$
      but with $|\Omega_c|\approx \sqrt{2\gamma_{32}/\gamma_{31}}\gamma$
      (and hence $x\ll 1$, $\sqrt{x}/z\sim1$), Im($K)_{\rm min}$ is small.
\end{itemize}

Similarly, ${\cal K}_j$ can be also expressed as the form (\ref{decom1}), with corresponding $\alpha_j$, $A_{j\pm}$, $\delta_{j+}$ and $\delta_{j-}$ given in the Appendix~\ref{app:co}. We decompose Im$({\cal K}_j)$ according to different $\Omega_c$ as the following.

(i). {\it Weak control field region}  (i.e. $\Omega_c<\Omega_{\rm ref}$): Im$({\cal K}_j)$
can be expressed as the form of Eq.~(\ref{eq:Form1_EIT}), but with
\begin{subequations}
\begin{eqnarray}
& &W_{1\pm}=\frac{1}{2}\left[\gamma_{31}(1+3B)+\gamma_{32}\pm
\sqrt{[\gamma_{31}(1-B)-\gamma_{32}]^2-4|\Omega_c|^2}\right],\\
& &W_{2\pm}=\frac{1}{2}\left[\gamma_{31}+\gamma_{32}+3\Delta\omega_D\pm\sqrt{(\gamma_{31}
-\gamma_{32}-\Delta\omega_D)^2-4|\Omega_c|^2}\right],\\
& &\Gamma^w_1=\gamma_{32}+2B\gamma_{31}+\Gamma_3|\Omega_c|^2(1-B)/[\Gamma_3(1-B^2)\gamma_{31}+2|\Omega_c|^2],\\
& &\Gamma^w_2=\gamma_{32}+2\Delta\omega_D+\Gamma_3|\Omega_c|^2(\gamma_{31}
-\Delta\omega_D)/[\Gamma_3(\gamma_{31}^2-\Delta\omega_D^2)+2\gamma_{31}|\Omega_c|^2].
\end{eqnarray}
\end{subequations}

\begin{figure}
\includegraphics[scale=0.75]{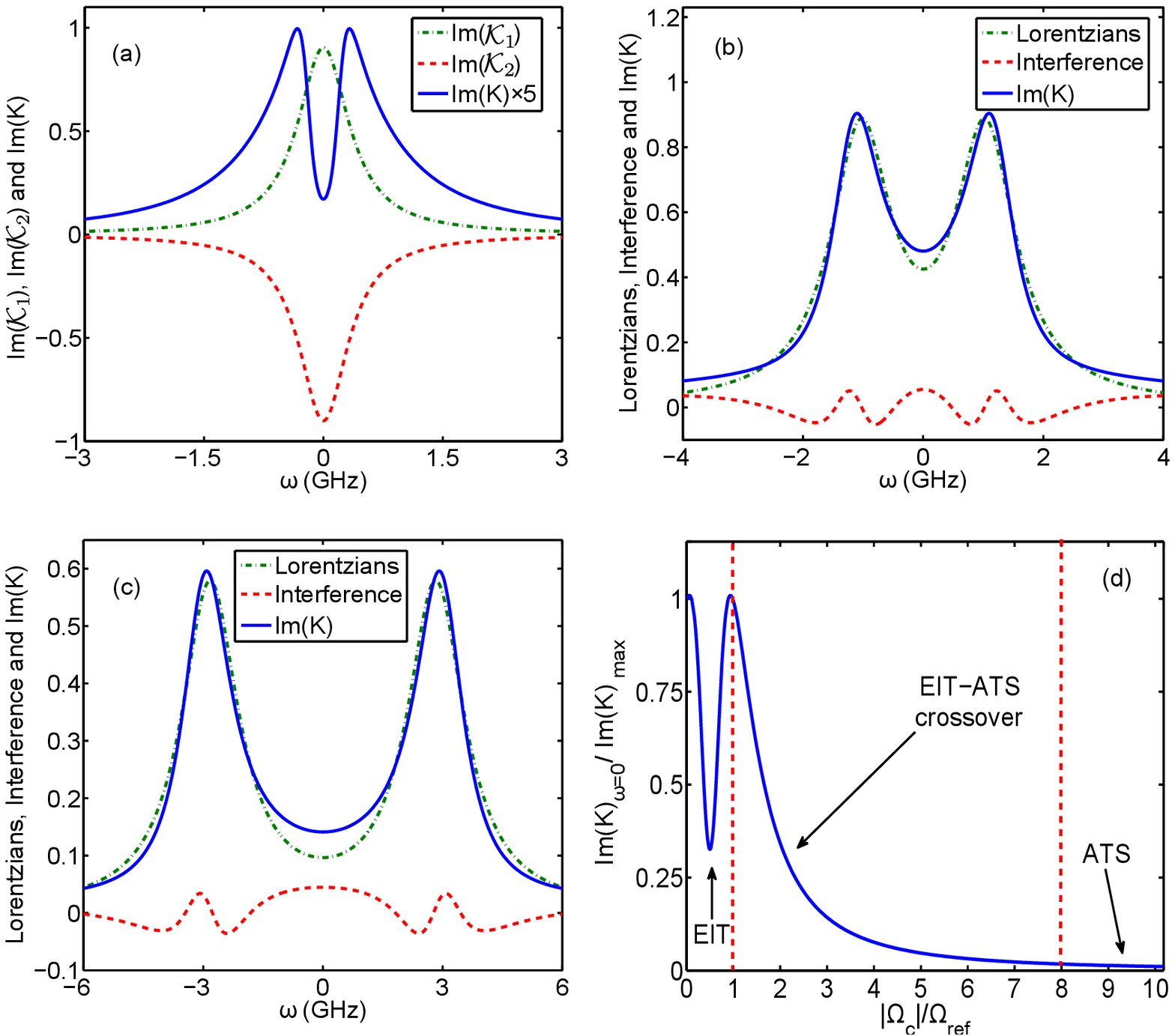}
\caption{(Color online) EIT-ATS crossover for hot molecules in the
counter-propagating configuration. (a):  Im(${\cal K}_1$)
(dash-dotted line), Im(${\cal K}_2$) (dashed line), and  ${\rm
Im}({K})$ (solid line) as function of $\omega$ for
$\Omega_c<\Omega_{\rm ref}$. (b): Absorption spectrum contributed by
two Lorentzians (dash-dotted line), the destructive interference
(dashed line), and Im($K$) (solid line), in the region
$\Omega_c>\Omega_{\rm ref}$. (c): Absorption spectrum contributed by
two Lorentzians (dash-dotted line), the destructive interference
(dashed line), and Im($K$) (solid line), in the region
$\Omega_c\gg\Omega_{\rm ref}$. Panels (a), (b) and (c) correspond to
EIT, EIT-ATS crossover, and ATS regions, respectively. (d):
Transition from EIT to ATS for hot molecules in the
counter-propagating configuration. ${\rm Im}(K)_{\omega=0}/{\rm
Im}(K)_{\rm max}$ as a  function of  $|\Omega_c|/\Omega_{\rm ref}$.
Three regions (EIT, EIT-ATS crossover, and ATS) are divided by two
dashed lines.} \label{L_D_counter}
\end{figure}

Shown in Fig.~\ref{L_D_counter}(a)
are results of Im(${\cal K}_1$) (the dash-dotted line), Im(${\cal K}_2$) (the dashed line),
and  Im(${K}$) (the solid line). System parameters are given by $\Gamma_{j2}
\approx \Gamma_{j3}\,\,(j=1,4)=4.08\times10^7$ Hz, $\gamma_{32}^{\rm col}\approx \gamma_{21}^{\rm col}\approx \gamma_{31}^{\rm col}= 5\times10^6$ Hz, $\gamma=3\times10^6$ Hz, $\Omega_c=100$ MHz, and $\Delta \omega_D=0.6$ GHz. Again, one has Im(${\cal K}_1)>0$ and Im(${\cal K}_2)<0$. Their sum gives total absorption Im($K$), which displays an absorption doublet with a significant transparency window near at $\omega=0$. This remarkable feature comes also from the destructive interference induced by the Doppler effect because the negative Im(${\cal K}_2)$ term disappears if the Doppler broadening is absent. According to the criterion given in Refs.~\cite{Petr2008,Tony2010,Petr2011}, such phenomenon belongs to EIT.  However, the counter-propagating configuration results in a mismatch of beam detunings relative to each other, and hence though a transparency window due to the Doppler effect is opened but it is relatively shallow comparing with the case of co-propagating configuration.

(ii). {\it Large control field region} (i.e. $\Omega_c>\Omega_{\rm ref}$):
By extending the approach by Agarwal \cite{Agarwal1997},
we obtain Im$({\cal K}_j)$ ($j=1,2$) with the same form of
Eq.~(\ref{eq:Form2_EIT}), but with
\begin{subequations}\label{eq:A1_counter2}
\begin{eqnarray}
& &W_1=(\gamma_{31}+\gamma_{32}+3\Delta\omega_D)/2,\\
& &\delta_1^r=\sqrt{4|\Omega_c|^2-(\gamma_{31}-\gamma_{32}-\Delta\omega_D)^2}/2,\\
& &g_1=\frac{\gamma_{31}-\gamma_{32}-\Delta\omega_D}{2}+\frac{\Gamma_3|\Omega_c|^2(\gamma_{31}
-\Delta\omega_D)}{\Gamma_3(\gamma_{31}^2-\Delta\omega_D^2)+2\gamma_{31}|\Omega_c|^2},
\end{eqnarray}
\end{subequations}
and
\begin{subequations}\label{eq:A2_counter2}
\begin{eqnarray}
& &W_2=[\gamma_{31}(1+3B)+\gamma_{32}]/2,\\
& &\delta_2^r=\sqrt{4|\Omega_c|^2-[\gamma_{31}(1-B)-\gamma_{32}]^2}/2,\\
& &g_2=\frac{\gamma_{31}(1-B)-\gamma_{32}}{2}+\frac{\Gamma_3|\Omega_c|^2(1-B)}{\Gamma_3(1-B^2)
\gamma_{31}+2|\Omega_c|^2}.
\end{eqnarray}
\end{subequations}

(iii) {\it Strong control field region} (i.e. $\Omega_c\gg\Omega_{\rm ref}$): In this situation, the quantum interference strength $g_j/\delta_j^r$ in the decomposed probe absorption spectrum (with the same form of Eq.~(\ref{eq:Form2_EIT})\,) is very weak and the linear absorption corresponds to the sum of two Lorentzians shifted from the origin by $\delta_j^r\approx\pm\Omega_c$ ($j=1,2$).

Shown in Fig.~\ref{L_D_counter}(b) and (c) are results of the probe-field absorption spectra as functions of $\omega$ for $\Omega_c>\Omega_{\rm ref}$ and $\Omega_c\gg\Omega_{\rm ref}$, respectively. The dash-dotted line (dashed line) denotes the contribution by the sum of two Lorentzians terms (interference terms) in Im($K$). We see that both destructive and constructive interferences appear for different $\omega$. The solid line gives the result of Im($K$). System parameters used are the same as those in panel (a) but with $\Omega_c=1.2$ and  GHz and $\Omega_c=3.0$ GHz for the panel (b) and the panel (c), respectively.

Shown in Fig.~\ref{L_D_counter} (d)
%
%
is the  ``phase diagram'' that
illustrates the transition from the EIT to ATS  for the conter-propagating configuration
by plotting ${\rm Im}(K)_{\omega=0}/{\rm Im}(K)_{\rm max}$ as a function of $|\Omega_c|/\Omega_{\rm ref}$.

\begin{figure}
\includegraphics[scale=0.4]{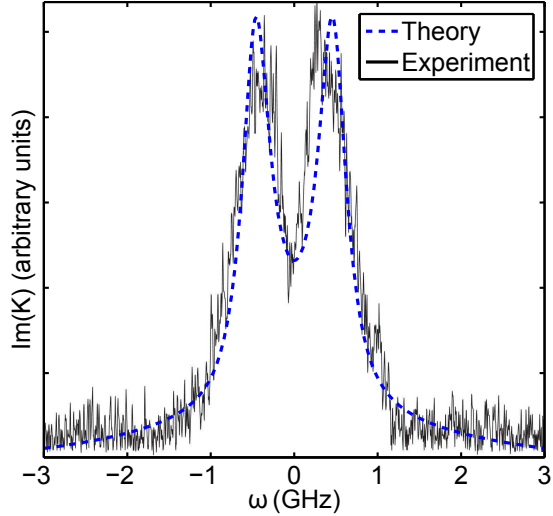}\\
\caption{(Color online) The experimental result  reported by  Lazoudis {\it et al}  \cite{Lazoudis2011}
(solid line) and our theoretical one (dashed line)  for the counter-propagating configuration. The control field is on resonance with $\Omega_c=240$ MHz.}\label{V_counter}
\end{figure}

To test our theoretical prediction, a comparison with the experimental one for the counter-propagating configuration reported by Lazoudis {\it et al} \cite{Lazoudis2011} is also made, as shown in
Fig.~\ref{V_counter}.
%
%
By using the system parameters $\Gamma_{12}=\Gamma_{13}=\Gamma_{42}=\Gamma_{43}=4.08\times10^7$ Hz, $\gamma_{jl}^{\rm col}=5\times10^6$ Hz, $\gamma=3\times10^6$ Hz, and $\Omega_c=240$ MHz,
the absorption spectrum  ${\rm Im}(K)$ is calculated based on our formulas, with the result plotted as the dashed line in the figure. We see that our theoretical result (the dashed line) agrees well with the experimental one measured by Lazoudis {\it et al} \cite{Lazoudis2011}, which is denoted by the solid line.

\section{Cold molecules and comparison for various cases}\label{Sec:cold_m}

Our model presented in Sec.~\ref{Sec:2} is also valid for cold molecules. In this case,
one should take $v=0$ in the Bloch Eq.~(\ref{dme}) and $f(v)=\delta (v)$ in the Maxwell
Eq.~(\ref{eqs:maxwell}). Solutions (\ref{eqs:init}) and (\ref{1osolution}) are still valid but
one must take $v=0$ and the dispersion relation is changed by
\begin{equation}\label{eq:K_cold_m}
K(\omega)=\frac{\omega}{c}+\kappa_{13}\sigma_{11}^{(0)}\frac{\omega+i\Gamma }{|\Omega_c|^2-(\omega+i\gamma_{31})(\omega+i\gamma_{32})},
\end{equation}
with $\Gamma=\gamma_{32}-(\Gamma_{12}|\Omega_c|^2)/(\Gamma_{12}\gamma_{21}+2|\Omega_c|^2)$.
We have chosen $\Delta_2=\Delta_3=0$ for simplicity.

The dispersion relation (\ref{eq:K_cold_m}) can also be decomposed as the form of Eq.~(\ref{decom1}),
but with spectrum poles given by
\begin{equation}\label{eq:d_pm}
\delta_{\pm}=[i(\gamma_{31}+\gamma_{32})\pm\sqrt{4|\Omega_c|^2-(\gamma_{31}-\gamma_{32})^2}]/2.
\end{equation}

\begin{figure}
  \includegraphics[scale=0.35]{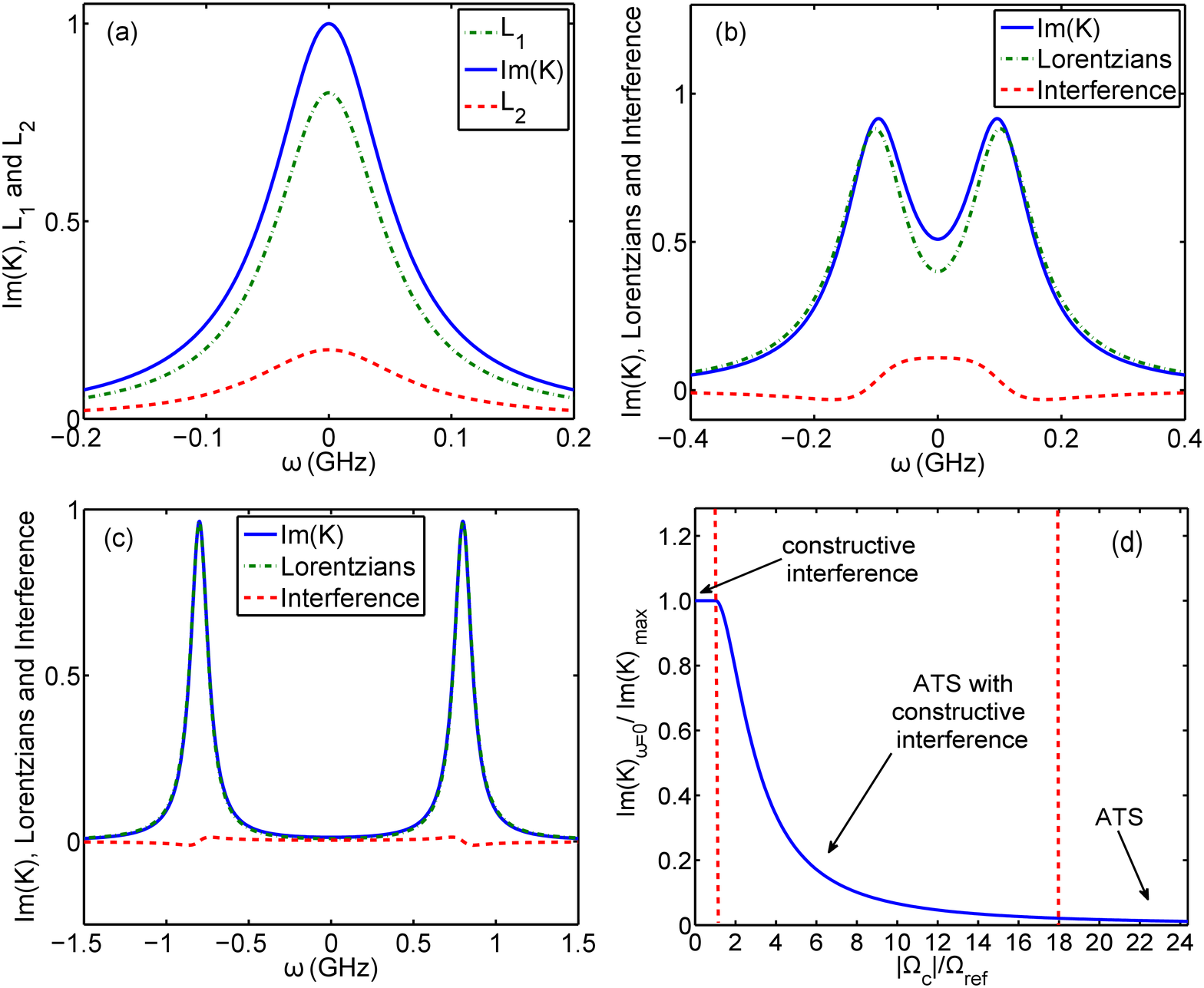}\\
  \caption{(Color online) (a): A single Lorentzian peak $L_1$ (dash-dotted line), a single Lorentzian peak $L_2$ (dashed line), and  ${\rm Im}({K})$ (solid line) as function of $\omega$ for $\Omega_c<\Omega_{\rm ref}$.
(b): Absorption spectrum contributed by two Lorentzians (dash-dotted
line), the constructive interference (dashed line), and Im($K$)
(solid line), in the region $\Omega_c>\Omega_{\rm ref}$. (c):
Absorption spectrum contributed by two Lorentzians (dash-dotted
line), the constructive interference (dashed line), and Im($K$)
(solid line), in the region $\Omega_c\gg\Omega_{\rm ref}$. (d):
${\rm Im}(K)_{\omega=0}/{\rm Im}(K)_{\rm max}$ for cold molecules as
the function of the control field $|\Omega_c|/\Omega_{\rm ref}$,
with $\Omega_{\rm ref}\equiv (\gamma_{32}-\gamma_{31})/2$. Three
regions (constructive interference, ATS with constructive
interference and ATS) are divided by two dashed
lines.}\label{fig;cold_cat}
\end{figure}

Similar spectrum decomposition can be done like that did in the last
section, which is omitted here for saving space.  Shown in panels
(a), (b) and (c) of Fig.~\ref{fig;cold_cat}
%
%
are the results of the probe-field absorption spectrum Im($K$)
decomposed in weak, large and strong control field region, respectively. System
parameters are given by
$\Gamma_{12}=\Gamma_{13}=\Gamma_{42}=\Gamma_{43}=4.08\times10^7$ Hz,
$\gamma_{jl}^{\rm col}=5\times10^6$ Hz, and $\sigma_{11}^{\rm eq}=1$ for
weak ($\Omega_c=18$ MHz; panel (a)), large ($\Omega_c=100$ MHz; panel (b)), and
strong ($\Omega_c=800$ MHz; panel (c)) control-field regions, respectively.
We see that in panel (a) the absorption spectrum is superposed by two
positive single Lorentzian peaks (i.e. the dashed line and the dotted-dashed
line), the superposition of them is a single peak (i.e. the solid line).
Thus the quantum interference is constructive
and hence no EIT in this weak control-field region. Similarly,
the quantum interferences shown in the panels (b) and (c) also constructive.
Consequently, there is no EIT and EIT-ATS crossover
for cold molecules in the V-type systems. Fig.~\ref{fig;cold_cat}(d) shows
${\rm Im}(K)_{\omega=0}/{\rm Im}(K)_{\rm max}$ for cold molecules as
a function of $|\Omega_c|/\Omega_{\rm ref}$,
where $\Omega_{\rm ref}\equiv (\gamma_{32}-\gamma_{31})/2$. Three
regions (constructive interference, ATS with constructive
interference and ATS) are divided by two vertical dashed
lines.


From these results and those given in section \ref{Sec:hot_m}, we see that the quantum interference in the V-type molecular system displays very different features, which depend on the existence or non-existence of the Doppler broadening, and also depend on the beam propagating
(co-propagating or counter-propagating) configurations. For comparison, in Table~\ref{table:EIT}
some useful physical quantities, including EIT condition, value ${\rm Im}(K)_{\omega=0}$,
group velocity $v_{g}$, and width of transparency window  $\Gamma_{TW}$ are presented for several
different physical cases.
%
\begin{table}
\caption{Comparison of propagating properties of the probe field for various V-type molecular systems, including EIT condition, value Im$(K)_{\omega=0}$, width of transparency window $\Gamma_{\rm TW}$, and group velocity $v_g$ for three different cases. Other quantities appeared in the Table have been defined in the text. Mol.=Molecules, Co-prop.=Co-propagating configuration, Cou.-prop.=Counter-propagating configuration.\label{table:EIT}}
\begin{ruledtabular}
\begin{tabular}{lllll}
 System & EIT condition &  Im$(K)_{\omega=0}$& $\Gamma_{\rm TW}$ & $v_g$ \\
  \hline
  Cold Mol. & no EIT &$ \frac{\kappa_{13}\gamma_{31}}{|\Omega_c|^2}$ & $\frac{2|\Omega_c|^2}{\gamma_{31}}$ & $\frac{|\Omega_c|^2}{\kappa_{13}}$\\
 Hot Mol. (Co-prop.) & $\gamma_{31}\gamma_{32}\leq|\Omega_c|^2\leq (\Delta\omega_D)^2/4$ & $\frac{\sqrt{\pi}\kappa_{13}}{\Delta\omega_D}\frac{\gamma_{32}\Gamma_{3}}{\gamma_{32}\Gamma_{3}
 +|\Omega_c|^2}$ & $\frac{2|\Omega_c|^2}{\Delta\omega_D}$ & $\frac{|\Omega_c|^2}{\sqrt{\pi}\kappa_{13}}$\\
 Hot Mol. (Cou.-prop.) & $2\gamma_{32}\gamma^2/\gamma_{31}\leq|\Omega_c|^2\leq(\Delta\omega_D)^2/4$ & $\frac{\sqrt{\pi}\kappa_{13}}{\Delta\omega_D}\frac{1}{1+|\Omega_c|/\Delta\omega_D}$ & $\frac{2|\Omega_c|^2}{\Delta\omega_D}$ & $\frac{|\Omega_c|^2}{\sqrt{\pi}\kappa_{13}}$\\

\end{tabular}
\end{ruledtabular}
\end{table}
%

The first line in Table~\ref{table:EIT} is for the cold molecular system, for which no EIT exists; the second line is for the Doppler-broadened system with the co-propagating configuration; the third line is for the Doppler-broadened system with the counter-propagating configuration. Both the co- and counter-propagating configurations allow Doppler-broadening-induced EIT, but their EIT conditions are different. The value of ${\rm Im}(K)_{\omega=0}$ for the co-propagating configuration is much less than those of the cold molecular system and the hot molecular system with the counter-propagating configuration. However, the width of transparency window and the group velocity  are the same for both  he co- and counter-propagating configurations. These interesting features deserve to be verified by
further experiments for V-type molecular systems.

\section{Roles of saturation and hole burning}\label{Sec:4}

Different from three-level $\Lambda$ system \cite{Lazoudis2010}, where only
quantum interference appears, in the three-level V-type system with inhomogeneous
broadening there may be a couple of simultaneously occurring
mechanisms (including saturation, hole burning, and quantum interference)
contributing to the absorption of the probe field,
which cannot be easily distinguished from each other. However,
the density matrix formulas are able to analyze all these various
contributions since they deal with both coherence and incoherence (population)
effects.

\begin{figure}
  \includegraphics[scale=0.4]{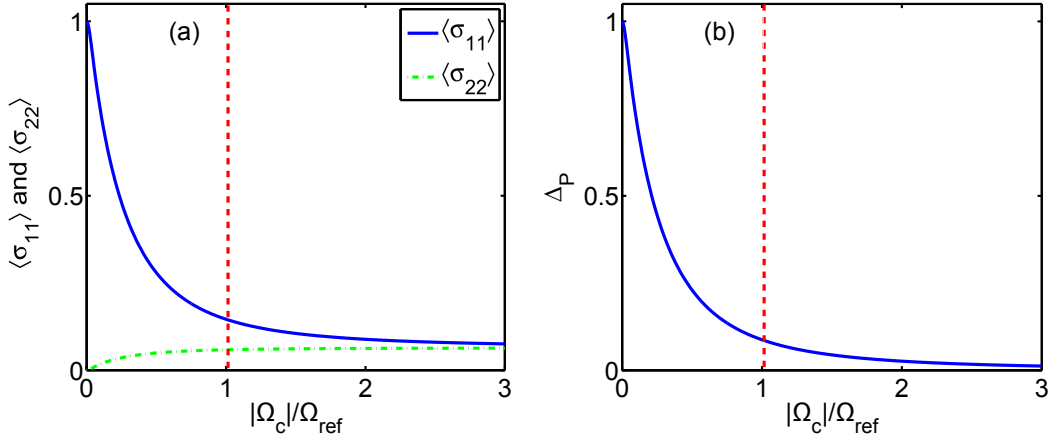}\\
  \caption{(Color online) (a) The average populations $\langle\sigma_{11}\rangle$ and $\langle\sigma_{22}\rangle$ in the states $|1\rangle$ (solid line) and $|2\rangle$ (dash-dotted line) as a function of $|\Omega_c|/|\Omega_{\rm ref}|$. (b) The average population difference $\Delta_{\rm P}$ as a function of $|\Omega_c|/\Omega_{\rm ref}$. The vertical dashed line in each
  panel is the boundary dividing weak (left side) and strong (right side) control-field regions.}\label{fig11}
\end{figure}

The saturation effect
is an incoherent phenomenon, which can be described by the difference between the
average populations in the states $|1\rangle$ and $|2\rangle$, i.e.
\begin{equation}\label{D}
\Delta_{\rm P}=\langle\sigma_{11}\rangle-\langle\sigma_{22}\rangle,
\end{equation}
where
$\langle\sigma_{jj}\rangle=\int_{-\infty}^{\infty}dvf(v)\sigma_{jj}$
 ($j=1,2)$, with $f(v)$ being velocity distribution function of atoms.
 If $\Delta_{\rm P}$ approaches zero, i.e.
$\langle\sigma_{11}\rangle\approx \langle\sigma_{22}\rangle$, the
system reaches maximum saturation. In contrary, if $\langle\sigma_{22}\rangle\ll \langle\sigma_{11}\rangle$ the saturation can be negligible.
For analyzing the saturation effect in our system, using the result
given in Sec.~\ref{Sec:2} we have calculated
$\langle\sigma_{11}\rangle$ and $\langle\sigma_{22}\rangle$, which are
plotted in Fig.~\ref{fig11}(a).
%
%
We see that in the weak control-field region
(i.e. the left side of the vertical dashed line)  $\langle\sigma_{11}\rangle$
is much lager than $\langle\sigma_{22}\rangle$. Shown in Fig.~\ref{fig11}(b)
is the result of $\Delta_{\rm P}$. These results demonstrate that only for very
strong control field (i.e. $|\Omega_c|/\Omega_{\rm ref}\gg 1$) the saturation
effect is significant. In the weak control
field region the saturation effect can be neglected.

Optical hole burning is an incoherent phenomenon where a saturating field
burns a hole into the population distribution for
an inhomogeneous broadened medium, which is usually called the
Lamb (or Bennet) hole when reflected in the absorption spectrum
of a probe field \cite{Wei2006}. In our case, the control field is
coupled to the states $|1\rangle$ and $|2\rangle$, and
the population in the state $|1\rangle$ indeed decreases when
$\Omega_c$ increases, as shown in Fig.~\ref{fig11}(a). However,
this phenomenon cannot be recognized as a hole burning.
The reasons are the following.  Firstly,  although for
the transition $|1\rangle\leftrightarrow|3\rangle$ coupled by the
probe field the control field is formally equivalent to a saturating field,
such saturation field has a large detuning to the state $|3\rangle$.
As a result, even if there exists a hole burning effect,
this effect is negligibly weak.
Secondly, in the absorption spectrum of the probe field shown in
Fig.~\ref{L_D_co}, there appears no Lamb hole that can be
taken as a signature of optical hole burning \cite{Dong2000}. A simple analysis
shows that the main reason for the reduction of $\langle\sigma_{11}\rangle$ comes from the
effect by the transient rate $\gamma$, together with the population transfer induced
by the control field.

Different from the above two incoherent effects which may occur in two-level
systems, the EIT is a quantum interference phenomenon occurring in multi (at least three) level
systems. From the above discussions we see that in the weak control-field region
both the saturation and hole burning play no significant role. Thus the
main reason of the reduction of probe-field absorption in the weak control-field region
is due to another mechanism.
As shown clearly in the panel (a) of Fig.~\ref{L_D_co},
where the absorption spectrum of the probe-field consists of a positive
(i.e. the red dashed line) and a negative (i.e. the green dashed-dotted line) part,
the reduction at the center of probe-field absorption spectrum is caused by
destructive interference, a typical character of EIT.
Similar character occurs also in cold three-level $\Lambda$ systems as
demonstrated in Ref.~\cite{Petr2011}, where EIT happens in the weak control-field region
in the same way (i.e. Im($K$) consists of two Lorentzians; one of them is positive and the other one is negative). When $\Omega_c$ increases to a large value,
the saturation and hole burning effects begin to take roles. However, the
ATS effect comes into play for large  $\Omega_c$ and dominates over the
saturation and hole burning effects when $\Omega_c$ becomes strong.
Consequently, there exists indeed a crossover from EIT to ATS in the system.

\section{Summary}\label{Sec:5}

We have studied EIT and ATS in open V-type molecular systems. A systematic analytical approach has been developed on the probe-field absorption spectrum by using residue theorem and spectrum-decomposition method. We have found that EIT can occur and there
exists a transition from EIT to ATS for hot molecules. However, there is no EIT and thus no EIT-ATS crossover for cold molecules. Furthermore, we have demonstrated
that for hot molecules an EIT is possible even for a counter-propagating configuration. We have provided explicit formulas of EIT conditions and widths of transparency windows of the probe field when hot molecules with Doppler broadening work in co-propagating and counter-propagating configurations, respectively. Our theoretical result agrees well with the recent experimental one reported by Lazoudis {\it et al}. \cite{Lazoudis2011}. New theoretical predictions presented in this work are useful for guiding new experimental findings in
coherent molecular systems and may have promising practical applications in coherent molecular spectroscopy, precision measurement, and molecular quantum state control, and so on.

\begin{acknowledgments}

This work was supported by NSF-China under Grant Nos. 10874043 and 11174080, and
by the Chinese Education Ministry Reward for Excellent Doctors in Academics under Grant No. MXRZZ2010007.
\end{acknowledgments}

\appendix

\section{Expressions of $\alpha$, $A_{j\pm}$, and $\delta_{j\pm}$}\label{app:co}

(i). For co-propagating configuration,
\begin{subequations}\label{corpoles}
\begin{eqnarray}
& & \alpha_1=\kappa_{13}'\times\frac{\Gamma_3(1-B^2)\gamma_{31}
+2|\Omega_c|^2}{\gamma\Gamma_3B(\Delta\omega_D^2
-B^2\gamma_{31}^2)},\\
& & \alpha_2=\kappa_{13}'\times\frac{\Gamma_3(\gamma_{31}^2-\Delta\omega_D^2)
+2\gamma_{31}|\Omega_c|^2}{\gamma\Gamma_3(\gamma_{31}^2-\Delta\omega_D^2)
+2\gamma_{31}|\Omega_c|^2\Gamma_{13}},\\
& & A_{1\pm}=\pm\left\{\delta_{1\pm}+ i\left[\gamma_{32}-\frac{\Gamma_3|\Omega_c|^2(1+B)}{\Gamma_3(1-B^2)\gamma_{31}
+2|\Omega_c|^2}\right]\right\}/(\delta_{1-}-\delta_{1+}),\\
& & A_{2\pm}=\pm\left\{\delta_{2\pm}+ i\left[\gamma_{32}-\frac{\Gamma_3|\Omega_c|^2(\gamma_{31}
+\Delta\omega_D)}{\Gamma_3(\gamma_{31}^2-\Delta\omega_D^2)+2\gamma_{31}|\Omega_c|^2}\right]\right\}/(\delta_{2-}-\delta_{2+}),\\
& & \delta_{1\pm}=\frac{1}{2}\left\{i\left[\gamma_{31}(1+B)
+\gamma_{32}\right]\pm\sqrt{4|\Omega_c|^2-[\gamma_{31}(1+B)-\gamma_{32}]^2}\right\},\\
& & \delta_{2\pm}=\frac{1}{2}\left[i(\gamma_{31}+\Delta\omega_D
+\gamma_{32})\pm\sqrt{4|\Omega_c|^2-(\gamma_{31}+\Delta\omega_D-\gamma_{32})^2}\right].
\end{eqnarray}
\end{subequations}
%

(ii). For counter-propagating configuration,
\begin{subequations}\label{courpoles}
\begin{eqnarray}
& & \alpha_1=\kappa_{13}'\times\frac{\Delta\omega_D\Gamma_3(1-B^2)\gamma_{31}+2|\Omega_c|^2}{\gamma\Gamma_3B(\Delta\omega_D^2-B^2\gamma_{31}^2)},\\
& & \alpha_2=\kappa_{13}'\times\frac{(\gamma_{31}^2-\Delta\omega_D^2)+|\Omega_c|^2}{\gamma(\gamma_{31}^2-\Delta\omega_D^2)+|\Omega_c|^2\Gamma_{13}},\\
& & A_{1\pm}=\pm\left\{\delta_{1\pm}+ i\left[\gamma_{32}+2B\gamma_{31}+\frac{\Gamma_3|\Omega_c|^2(1-B)}{\Gamma_3(1-B^2)\gamma_{31}+2|\Omega_c|^2}\right]\right\}/(\delta_{1-}-\delta_{1+}),\\
& & A_{2\pm}=\pm\left\{\delta_{2\pm}+ i\left\{\gamma_{32}+2\Delta\omega_D+\frac{\Gamma_3|\Omega_c|^2(\gamma_{31}-\Delta\omega_D)}{\Gamma_3(\gamma_{31}^2-\Delta\omega_D^2)+2\gamma_{31}|\Omega_c|^2}\right]\right\}/(\delta_{2-}-\delta_{2+}),\\
& & \delta_{1\pm}=\frac{1}{2}\left\{i\left[\gamma_{31}(1+3B)+\gamma_{32}\right]
\pm\sqrt{4|\Omega_c|^2-[\gamma_{31}(1-B)-\gamma_{32}]^2}\right\},\\
& & \delta_{2\pm}=\frac{1}{2}\left[i(\gamma_{31}+\gamma_{32}
+3\Delta\omega_D)/\pm\sqrt{4|\Omega_c|^2-(\gamma_{31}-\gamma_{32}-\Delta\omega_D)^2}\right].
\end{eqnarray}
\end{subequations}
%



\begin{references}
\bibitem{note2}The frequency and wavevector of the probe field are given by
$\omega_p+\omega$ and $k_p+K(\omega)$, respectively. Thus $\omega=0$ corresponds to
the center frequency of the probe field.
\end{references}
\end{document}